\documentclass[12pt]{article}
\usepackage{eurosym,cite}
\usepackage[left=0.7in, right=0.7in, top=0.7in, bottom=0.7in]{geometry}
\usepackage{hyperref}
\usepackage{mathtools}
\usepackage{graphicx}
\usepackage{mathrsfs}
\usepackage[utf8]{inputenc}
\usepackage[T1]{fontenc}
\usepackage{mathrsfs}
\usepackage{subcaption}
\usepackage{braket}
\usepackage{tikz}
\usepackage{tikz-cd}
\usepackage{pgfplots}

\begin{document}
	
	\begin{center}
		\textbf{\LARGE Cavity magnomechanical framework for a high-efficiency quantum battery\\}
		\bigskip
		S. K. Singh$^{a}$, \textit{\footnote{e-mail:singhshailendra3@gmail.com}},
		Ahmed A. Zahia$^{b}$ \textit{\footnote{e-mail:ahmed.zahia@fsc.bu.edu.eg}},  Jia-Xin Peng$^{c}$ \textit{\footnote{e-mail:JiaXinPeng@ntu.edu.cn}},
		M.Y.Abd-Rabbou$^{d,e}$ \textit{\footnote{e-mail:m.elmalky@azhar.edu.eg}},  
		
		$^{a}${\footnotesize Department of Physics, Akal University, Talwandi Sabo, Bathinda, Punjab 151302, India.}\\
		$^{b}${Department of Mathematics, Faculty of Science, Benha University, Benha, Egypt}\\
		$^{c}${\footnotesize School of Physical Science and Technology, Nantong University, Nantong 226019, People’s Republic of China.}\\
		$^{d}${\footnotesize School of Physics, University of Chinese Academy of Sciences, Yuquan Road 19A,
		Beijing 100049, People’s Republic of China.}\\
		$^{e}${\footnotesize Mathematics Department, Faculty of Science, Al-Azhar University, Nasr City 11884, Cairo, Egypt.}\\	
		
	\end{center}

	\begin{abstract}
		We theoretically investigate a quantum battery architecture where two identical two-level atoms are charged by a  cavity-magnomechanical system, which includes a microwave cavity, a magnon mode hosted in a YIG sphere, and  phonon mode due to deformation of YIG sphere. The charging process relies on coherent energy exchange, where the atoms couple to the cavity, which in turn interacts with the magnon mode via a beam-splitter mechanism.By deriving the system Hamiltonian under the rotating-wave approximation and employing a Lindblad master equation to rigorously model dissipation, we analyze the complete dynamical evolution of the battery. Our study demonstrates that strong, resonant light-matter interactions are crucial for enhancing the key performance metrics; charging efficiency, stored energy, and ergotropy (extractable work). We systematically investigate the deleterious effects of detuning and decoherence, and critically, we uncover a non-trivial interplay between the system's coupling strengths. This reveals optimal operating regimes where constructive interference maximizes performance, while excessive coupling in specific channels can degrade it. Ultimately, our findings provide a quantitative framework for engineering high-efficiency quantum batteries in hybrid magnonic platforms, offering a design roadmap for future experimental realizations.		
\end{abstract}

	\section{Introduction}
	The ongoing development of quantum technologies has spurred the investigation of novel energy storage devices capable of satisfying stringent operational requirements for speed and low dissipation. In this context, quantum batteries (QBs), have emerged as a pivotal area of research. Unlike conventional electrochemical batteries~\cite{liu2014comparative,chaoui2017aging,chen2012state,fleischer2013adaptive,rivera2017soc}, QBs utilize quantum superposition, coherence, and entanglement to enhance charging efficiency, storage capacity, and power output~\cite{alicki2013entanglement,binder2015quantacell,campaioli2017enhancing,ghosh2020enhancement,farina2019charger,ghosh2022dimensional,zahia2025optimizing}. Recent studies have demonstrated that many-body quantum effects can provide significant advantages: for instance, the quantacell model reveals a collective $N$-fold charging power enhancement~\cite{binder2015quantacell}, while spin-chain and Dicke-type batteries exhibit cooperative speedups beyond classical limits~\cite{ferraro2018high,crescente2020ultrafast}.
	
	\par\vspace{\baselineskip}
	
	A wide range of physical platforms has been proposed for implementing QBs, including spin systems~\cite{yao2022optimal,abd2025charging,arjmandi2023localization}, atomic ensembles in optical cavities~\cite{andolina2019extractable,zhang2023quantum}, and superconducting circuits~\cite{dou2023superconducting}. While theoretical models confirm that charging power can scale with system size~\cite{binder2015quantacell,campaioli2017enhancing,julia2020bounds}, practical realizations are invariably confronted by a principal challenge: decoherence induced by environmental coupling, which leads to deleterious self-discharging. To counteract this degradation, a portfolio of sophisticated strategies has been proposed, including the exploitation of disordered local fields~\cite{arjmandi2022enhancing}, the use of dark states to stabilize energy storage~\cite{quach2020using}, and the leveraging of non-Markovian dynamics~\cite{kamin2020non,morrone2023charging}, each showing promise in prolonging the usable lifetime of QBs. Concurrently, a fundamental line of inquiry has focused on elucidating the role of quantum correlations in QB performance. While entanglement is not strictly necessary for optimal work extraction~\cite{alicki2013entanglement,andolina2019extractable,zahia2025entanglement}, it can enhance charging under specific protocols~\cite{santos2020stable,arjmandi2022performance,zahia2025optimizing}. Coherent quantum charging schemes have also been shown to outperform classical strategies, offering improved robustness against decoherence and energy loss~\cite{andolina2019extractable}. Collectively, these findings underscore the transformative potential of QBs as next-generation energy storage devices, capable of powering emerging quantum technologies with high efficiency, scalability, and resilience~\cite{yang2023battery,zahia2024quantum,gyhm2022quantum,francica2024quantum}.
	
	The magnon mode, a collective excitation of spins in magnetic materials, has attracted considerable interest due to its high spin density, low damping rates, and tunability. These properties enable robust coupling between magnon modes in YIG and microwave cavity photon modes~\cite{li2018magnon,zhang2016cavity,zahia2023bidirectional,khalil2025controlling}, with experiments having decisively confirmed the attainment of the strong coupling regime in YIG spheres~\cite{tabuchi2014hybridizing,zhang2014strongly,kumar2023reconfigurable}. Exploiting this robust interaction, hybrid ferrimagnetic systems have served as a fertile ground for demonstrating a range of quantum phenomena, including magnon dark modes and gradient memory~\cite{zhang2015magnon}, exceptional points~\cite{zhang2017observation}, bistability of cavity--magnon polaritons~\cite{wang2018bistability}, magnon cooling~\cite{sharma2018optical}, and the Kerr effect in magnons~\cite{wang2016magnon}. Recent advances also include milestones such as the single-shot detection of individual magnons~\cite{lachance2020entanglement}, the generation of quantum entanglement, and remote asymmetric quantum steering~\cite{hidki2024generation,cong2023entanglement,hidki2024asymmetric,xie2023stationary,zuo2025controllable}.
	
	\par\vspace{\baselineskip}
	
	Building upon this foundation, the field of cavity magnomechanics leverages the nonlinear effects of magnetostrictive interactions, linear magnetic dipolar coupling, Josephson parametric amplification, and magnon squeezing to create quantum correlations in diverse physical architectures~\cite{li2019entangling,lakhfif2024maximum,li2021entangling,hidki2024entanglement,amazioug2023feedback}. This approach, analogous to cavity optomechanics, offers a versatile paradigm to study new physical effects and apply them in quantum technologies~\cite{lachance2019hybrid}. In these systems, phenomena such as radiation pressure mediate the coupling between magnomechanical displacement and optical cavities~\cite{fan2022optical,fan2023entangling,fan2022stationary,fan2023microwave,tadesse2024distant,peng2025symmetric}, allowing entanglement to form between microwave and optical fields. Additionally, this class of interaction has been shown to be capable of producing entangled states involving atomic ensembles and magnon modes~\cite{kong2022magnon,dilawaiz2024entangled}.

\par\vspace{\baselineskip}

The primary motivation for this work is to conduct a systematic theoretical investigation into the influence of frequency detunings, coupling strengths, and environmental interactions on the performance of a quantum battery realized within a hybrid cavity–magnomechanical system. This platform allows for rich, coherent interactions among photonic, magnonic, and mechanical modes, providing an ideal setting to study quantum energy transfer and storage. Our analysis centers on the coherence of the charger subsystem and key performance benchmarks, such as stored energy and ergotropy, which quantifies the maximum extractable work. The dependence of these quantities on system parameters plays a key role in identifying optimal regimes for maximum efficiency, stability, and resilience. This analysis reveals the physical mechanisms governing energy exchange and retention, offering critical insights for the design of efficient, robust quantum batteries for future quantum technologies.

\par\vspace{\baselineskip}

The paper is organized as follows. In Section~\ref{sec1}, we present the Hamiltonian of the hybrid cavity-magnomechanical system and outline the derivation of the Schrödinger equation that governs its dynamics. Section~\ref{sec2} analyzes the coherence dynamics of the charger under variations of the model parameters. In Section~\ref{sec3}, we investigate the storage process and the stability of the quantum battery. Section~\ref{sec4} explores the maximum extractable work from the system, whereas Section~\ref{pur} discusses the degree of coherence and mixedness of the quantum states, providing insight into how close the system remains to an ideal pure state. Finally, Section~\ref{sec5} summarizes the main findings and highlights their implications for the design of efficient quantum batteries. 

	\section{The Model}\label{sec1}
\begin{figure}[h!]
		\centering
		\includegraphics[width=0.8\textwidth]{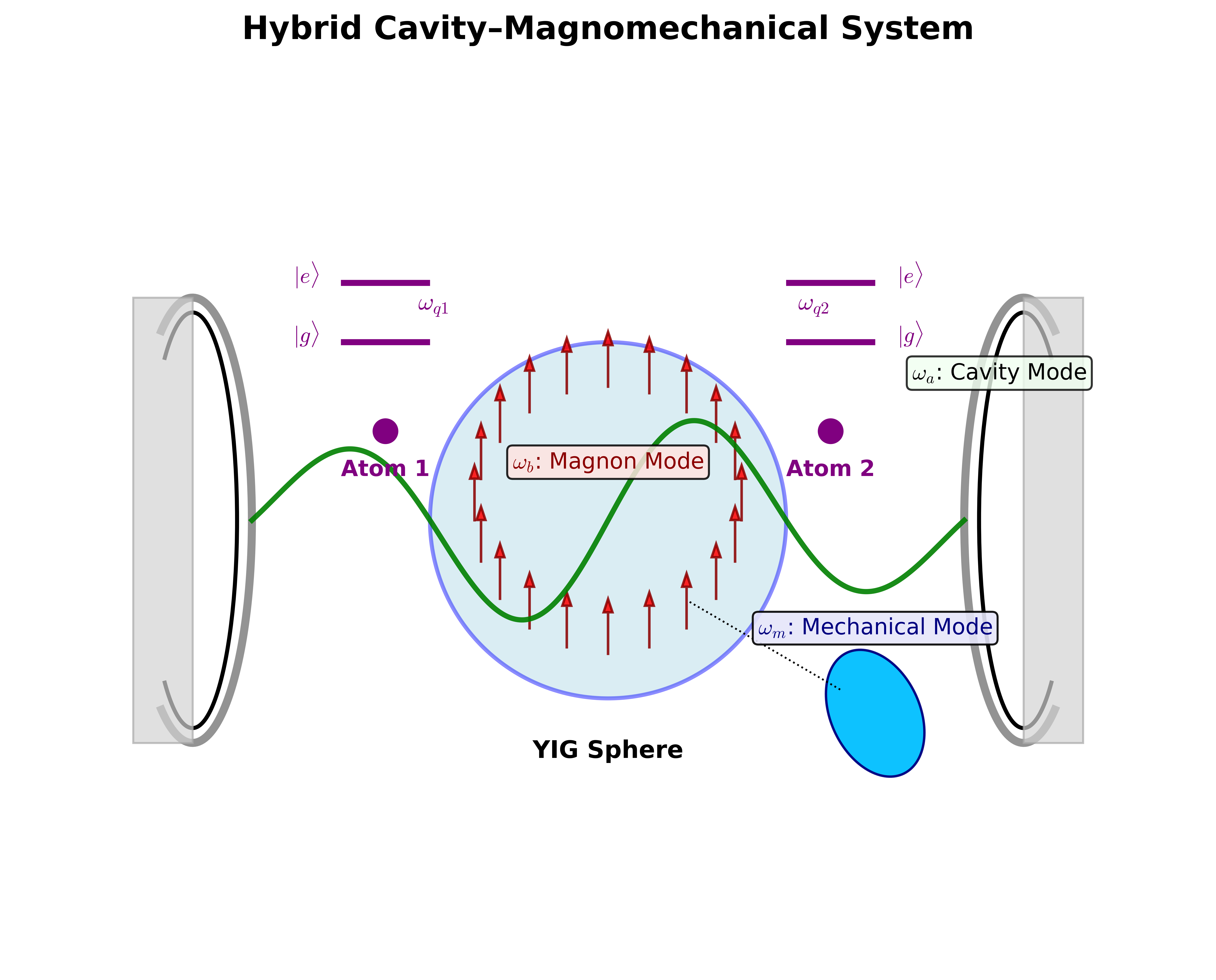}
		\caption{Schematic of a hybrid cavity magnomechanical quantum battery where two identical two-level atoms (frequency $\omega_q$) are coupled to the microwave cavity mode. Energy transfer and storage are mediated by cavity--magnon coupling $g_a$, phonon--magnon coupling $g_b$, and atom--cavity interaction $\lambda$, with respective decay rates $\kappa_{a,b,m}$ and $\gamma$.}
		\label{model}
	\end{figure}

In this section, we present the theoretical model of a hybrid cavity-magnomechanical system for a quantum battery, as illustrated in Fig.~\ref{model}. The setup incorporates two identical two-level atoms, which constitute the quantum battery. The charger is a hybrid system comprising a microwave cavity mode, a magnon mode hosted within a YIG-sphere, and a mechanical phonon mode of the sphere's vibration. The coupling between magnons and cavity photons occurs through the magnetic dipole interaction, whereas the magnetostrictive interaction mediates the coupling between the magnon mode and \textcolor{red}{the} phonon mode. This magnetostrictive coupling leads to the magnon-induced deformation of the YIG sphere’s geometric structure and the formation of phonon modes.
	 
	The total Hamiltonian of this hybrid quantum system is given by:
	\begin{align}
		\hat{H} =\ & \omega_a\, \hat{a}^\dagger \hat{a} 
		+ \omega_b\, \hat{b}^\dagger \hat{b} 
		+ \omega_m \left( \hat{q}^2 + \hat{p}^2 \right)
		+ g_b\, (\hat{b}^\dagger + \hat{b}) \hat{q}
		+ g_a\, (\hat{a}^\dagger \hat{b} + \hat{b}^\dagger \hat{a})\notag \\
		&+ \sum_{j=1}^{2} \left(
		\frac{\omega_{q}}{2}\, \hat{\sigma}_z^{(j)} 
		+ \lambda_{j}\, \hat{\sigma}_x^{(j)} (\hat{a} + \hat{a}^\dagger)
		- i \frac{\gamma}{2}\, \hat{\sigma}_+^{(j)} \hat{\sigma}_-^{(j)}
		\right) 
		- i \frac{\kappa_a}{2}\, \hat{a}^\dagger \hat{a}
		- i \frac{\kappa_b}{2}\, \hat{b}^\dagger \hat{b}
		- i \frac{\kappa_m}{2}\, \hat{m}^\dagger \hat{m}. \label{Ham}
	\end{align}

Here, $\hat{a}$ and $\hat{b}$ ($\hat{a}^\dagger$, $\hat{b}^\dagger$) are the annihilation (creation) operators for the microwave cavity and magnon mode, respectively. The dimensionless position and momentum operators of the phonon (mechanical) mode are defined as $\hat{q} = \frac{1}{\sqrt{2}} (\hat{m} + \hat{m}^\dagger)$ and $\hat{p} = \frac{i}{\sqrt{2}} (\hat{m}^\dagger - \hat{m})$. The two-level atoms are described by the Pauli spin operators, with $\hat{\sigma}_+ = |e\rangle \langle g|$ and $\hat{\sigma}_- = |g\rangle \langle e|$ being the raising and lowering operators, respectively. From these, we define $\hat{\sigma}_x = \hat{\sigma}_+ + \hat{\sigma}_-$ and $\hat{\sigma}_z = [\hat{\sigma}_+, \hat{\sigma}_-]$. The parameters $\omega_a$ and $\omega_b$ denote the frequency of the microwave cavity and magnon mode, respectively, whereas $\omega_m$ is the phonon frequency and $\omega_{q}$ is the transition frequency of the two-level atoms. The coupling strength between the microwave cavity and the magnon mode is $g_{a}$, whereas $g_{b}$ denotes the coupling between the magnon and phonon modes. Each atom couples to the cavity mode with strength $\lambda_j$ ($j=1,2$). The atomic spontaneous emission rate is given by $\gamma$, whereas $\kappa_a$, $\kappa_b$, and $\kappa_m$ are the decay rates of the cavity, magnon, and phonon modes, respectively.

The Hamiltonian in Eq.~\eqref{Ham} is manifestly non-Hermitian, incorporating imaginary terms that phenomenologically describe dissipation. This effective Hamiltonian approach provides a computationally efficient method to model system dynamics under the assumption of a zero-temperature Markovian environment~\cite{gardiner2004quantum,el2018non}. Specifically, this formalism is valid in a regime where the evolution of the system's state vector is conditioned on the absence of a "quantum jump" event (e.g., the emission of a photon into the environment). It correctly captures the continuous decay of probability amplitudes within a given excitation subspace due to irreversible energy loss, but it does not account for thermal repopulation or stochastic back-action from the environment. Given that our analysis is confined to the single-excitation subspace and we are primarily interested in the coherent transfer dynamics and subsequent decay, this method offers a well-established and insightful approximation of the full open-system evolution~\cite{plenio1998quantum}.

On applying the rotating wave approximation  (RWA) and transforming into the interaction picture, the effective interaction Hamiltonian of the system is given by \cite{zeuch2020exact}:
	\begin{align}
		\hat{H}_I =\ & g_b\, \left( \hat{b}^\dagger \hat{m}\, e^{i\delta_1 t} + \hat{b} \hat{m}^\dagger\, e^{-i\delta_1 t} \right) 
		+ g_a\, \left( \hat{a}^\dagger \hat{b}\, e^{i\delta_2 t} + \hat{a} \hat{b}^\dagger\, e^{-i\delta_2 t} \right)\notag \\
		&+ \sum_{j=1}^{2} \left[
		\lambda_{j}\,  \left( \hat{\sigma}_+^{(j)} \hat{a}\, e^{i\delta_3 t} + \hat{\sigma}_-^{(j)} \hat{a}^\dagger\, e^{-i\delta_3 t} \right)
		- i \frac{\gamma}{2}\, \hat{\sigma}_+^{(j)} \hat{\sigma}_-^{(j)}
		\right] 
		- i \frac{\kappa_a}{2}\, \hat{a}^\dagger \hat{a}
		- i \frac{\kappa_b}{2}\, \hat{b}^\dagger \hat{b}
		- i \frac{\kappa_m}{2}\, \hat{m}^\dagger \hat{m}
	\end{align}
	
	Here, the detuning parameters are defined as $\delta_1 = \omega_b - \omega_m$, $\delta_2 = \omega_a - \omega_b$, and $\delta_3 = \omega_q - \omega_a$, representing the frequency differences between the interacting modes. The initial state of the total system is chosen as
	\[
	\ket{\psi(0)} = \ket{g g} \otimes \ket{1\,0\,0},
	\]
where both two-level atoms are in their ground state $\ket{g}$, while the charger begins with a single excitation in the cavity mode ($\ket{1}_{\text{cav}}$), and vacuum states for the magnon ($\ket{0}_{\text{mag}}$) and phonon ($\ket{0}_{\text{ph}}$) modes.

Let the time-dependent state of the system be expressed as:
	\begin{align}
		\ket{\psi(t)} =\ & C_1(t)\ket{g g} \otimes \ket{1\,0\,0}
		+ C_2(t)\ket{g g} \otimes \ket{0\,1\,0}
		+ C_3(t)\ket{g g} \otimes \ket{0\,0\,1} \notag \\
		&+ C_4(t)\ket{e g} \otimes \ket{0\,0\,0}
		+ C_5(t)\ket{g e} \otimes \ket{0\,0\,0}
	\end{align}
	
	By applying the time-dependent Schrödinger equation
	
	$i\frac{d}{dt} \ket{\psi(t)} =  \hat{H}_I \ket{\psi(t)}$,
	we obtain the following system of coupled differential equations for the probability amplitudes:
\begin{align}
	i \dot{C}_1(t) &= -i \frac{\kappa_a}{2} C_1(t) + g_a e^{-i \delta_2 t} C_2(t) + 2\lambda e^{-i \delta_3 t} C_4(t), \notag \\
	i \dot{C}_2(t) &= g_a e^{i \delta_2 t} C_1(t) - i \frac{\kappa_b}{2} C_2(t) + g_b e^{-i \delta_1 t} C_3(t), \notag \\
	i \dot{C}_3(t) &= g_b e^{i \delta_1 t} C_2(t) - i \frac{\kappa_m}{2} C_3(t), \label{eq:diffeqsystem} \\
	i \dot{C}_4(t) &= \lambda e^{i \delta_3 t} C_1(t) - i \frac{\gamma}{2} C_4(t). \notag
\end{align}
	
	Due to the symmetry between the two qubits, we have $C_5(t) = C_4(t)$, and $\lambda_j=\lambda$ for $j\in\{1,2\}$. Since the system of equations contains time-dependent coefficients, we apply the transformation
	\[
	C_n(t) = Z_n(t)\, e^{-i \Delta_n t}, \quad \text{for } n \in \{1, 2, 3, 4\},
	\]
	in order to eliminate the explicit time-dependence in the exponential terms, where $\Delta_n$ are appropriately chosen reference frequencies for each amplitude.
	
	Now, the whole system of differential equations are 
	\begin{align}
		i \dot{Z}_1(t) &=y_1Z_1(t)+g_aZ_2(t)+2\lambda Z_4(t),\notag\\
		i \dot{Z}_2(t) &=g_aZ_1(t)+y_2Z_2(t)+g_b Z_3(t),\notag\\
		i \dot{Z}_3(t) &=g_bZ_2(t)+y_3Z_3(t),\\
		i \dot{Z}_4(t) &=\lambda Z_1(t)+y_4Z_4(t)\notag.
	\end{align}
	where 
	\begin{align}
		y_1 &=-(i\frac{\kappa_a}{2}+\Delta_1), \qquad \Delta_1=2\delta_3-\delta_2,\notag\\
		y_2 &=-(i\frac{\kappa_b}{2}+\Delta_2), \qquad \Delta_2=2\delta_3-2\delta_2,\notag\\
		y_3 &=-(i\frac{\kappa_m}{2}+\Delta_3), \qquad \Delta_3=2\delta_3-2\delta_2-\delta_1,\\
		y_4 &=-(i\frac{\gamma}{2}+\Delta_4), \qquad \Delta_4=\delta_3-\delta_2,\notag
	\end{align}
	The system is expressed as $i\frac{d}{dt}\mathcal{Z}(t)=\mathcal{A}\mathcal{Z}(t)$, where 
	\begin{equation}
		\mathcal{Z}(t)=\begin{bmatrix}
			Z_1(t) \\
			Z_2(t) \\
			Z_3(t) \\
			Z_4(t)
		\end{bmatrix}, \quad \mathcal{A}=\begin{bmatrix}
			y_1 & g_a & 0 & 2\lambda \\
			g_a & y_2 & g_b & 0 \\
			0 & g_b & y_3 & 0 \\
			\lambda & 0 & 0 & y_4
		\end{bmatrix}.
	\end{equation}
	Therefore, the general solution of the system of differential equations can be formulated as
	\begin{equation}
		\mathcal{Z}(t) = \exp\left(-i \mathcal{A}t \right)\, \mathcal{Z} (0). \tag{13}
	\end{equation}
	
	This general solution can be reformulated based on the eigenvalues $\varphi_j$ and eigenstates $\ket{\phi_j}$ as
	\begin{equation}
		\mathcal{Z}(t) = \sum_{j=1}^{4} \exp(-i \varphi_j t)\ket{\phi_j}\bra{\phi_j}\mathcal{Z}(0). \tag{14}
	\end{equation}
	
	where the exponential of matrix $\exp\left(-i \mathcal{A}t\right)$ or the eigenvalues $\varphi_j$ and corresponding eigenstates $\ket{\phi_j}$ can be calculated by Python.
	
	Now, since the total density operator is given by $\hat{\rho}(t) = \ket{\psi(t)}\bra{\psi(t)}$, we can obtain the reduced density matrix of the atomic system (denoted as QB) by performing a partial trace over the field degrees of freedom, i.e., $\mathrm{Tr}_F[\hat{\rho}(t)]$. This yields:
	\begin{align}
		\hat{\rho}_{\mathrm{QB}}(t) =\ & \left(|C_1(t)|^2 + |C_2(t)|^2 + |C_3(t)|^2\right) \ket{gg}\bra{gg}
		+ |C_4(t)|^2 \ket{eg}\bra{eg}
		+ |C_5(t)|^2 \ket{ge}\bra{ge} \notag \\
		& + C_4(t)C_5^*(t) \ket{ge}\bra{eg}
		+ C_4^*(t)C_5(t) \ket{eg}\bra{ge}.
	\end{align}
	
	The same procedure can be applied to obtain the reduced density matrix of the charger (field), denoted by $\hat{\rho}_F(t)$, by taking the partial trace over the atomic system degrees of freedom, i.e., $\mathrm{Tr}_{\mathrm{QB}}[\hat{\rho}(t)]$. This gives:
	\begin{align}
		\hat{\rho}_F(t) =\ & |C_1(t)|^2 \ket{100}\bra{100}
		+ |C_2(t)|^2 \ket{010}\bra{010}
		+ |C_3(t)|^2 \ket{001}\bra{001} \notag \\
		& + \left(|C_4(t)|^2 + |C_5(t)|^2\right) \ket{000}\bra{000} \notag \\
		& + C_1(t)C_2^*(t) \ket{100}\bra{010}
		+ C_1^*(t)C_2(t) \ket{010}\bra{100} \notag \\
		& + C_1(t)C_3^*(t) \ket{100}\bra{001}
		+ C_1^*(t)C_3(t) \ket{001}\bra{100} \\
		& + C_2(t)C_3^*(t) \ket{010}\bra{001}
		+ C_2^*(t)C_3(t) \ket{001}\bra{010}. \notag
	\end{align}
	
	\section{Coherence}\label{sec2}
We begin our analysis by investigating the quantum coherence within the charger subsystem, which comprises the microwave cavity, magnon, and phonon modes. Coherence is a fundamental quantum resource that quantifies the ability of a system to exist in a superposition of its basis states; in this context, it reflects the integrity of the phase relationships crucial for coherent energy transfer. A standard, basis-dependent measure of coherence is the $l_1$-norm, defined as the sum of the absolute values of the off-diagonal elements of the density matrix~\cite{cimini2019measuring}:
	\begin{align}
		C_{l_1}(\rho) &= \sum_{i \neq j} |\rho_{ij}|\notag\\
		&=2|C_1(t)C_2(t)|+2|C_1(t)C_3(t)|+2|C_2(t)C_3(t)|
	\end{align} 
	where $\rho$ is the density matrix, and $\rho_{ij}$ are its off-diagonal elements representing superpositions.
	
	\begin{figure}[h!]
		\centering
		\includegraphics[width=1.05\textwidth]{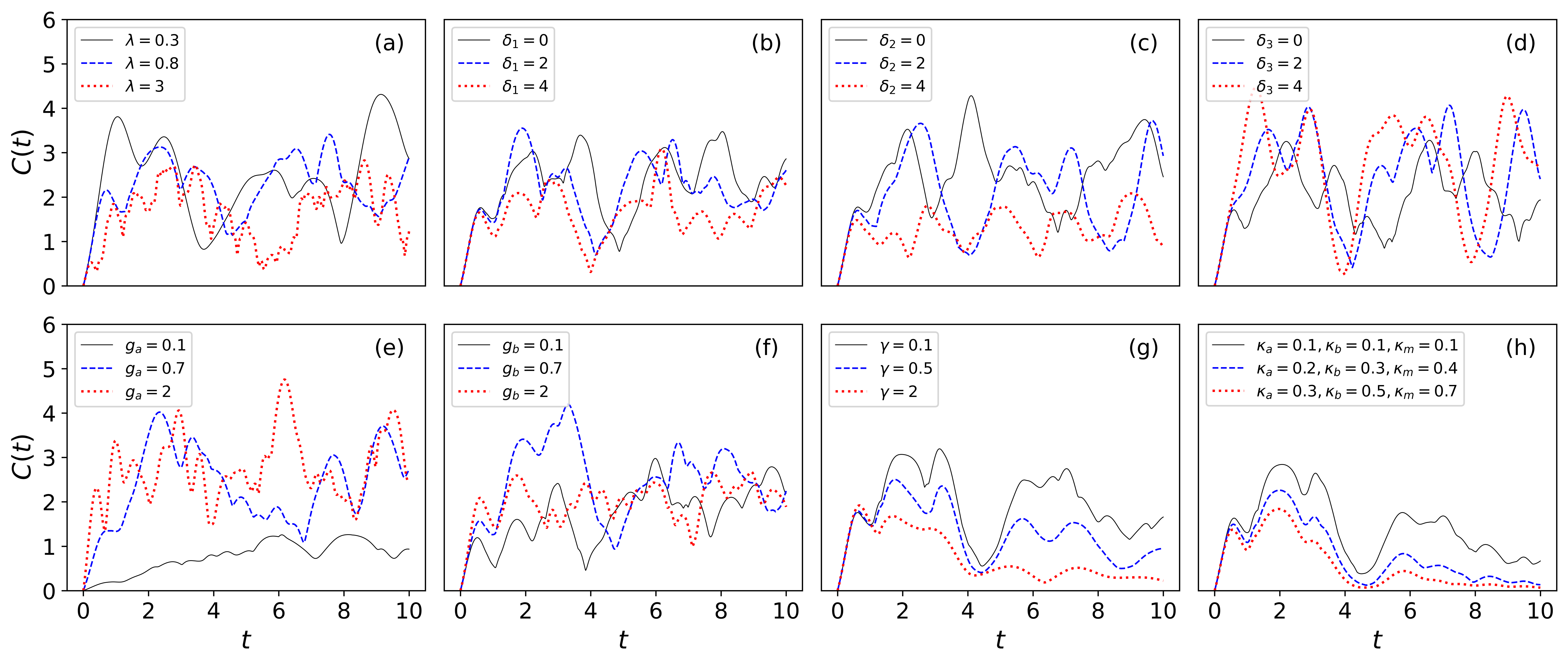}
		\caption{Coherence dynamics $C(t)$ of the charger as a function of time $t$. 
			(a) $\delta_i = 1$ for $i = 1, 2, 3$, $g_a = g_b = 1$, $\kappa_a = \kappa_b = \kappa_m = \gamma = 0$, with varying $\lambda$. 
			(b)  $\delta_2 = \delta_3 = 1$, $g_a = g_b = 1$, $\kappa_a = \kappa_b = \kappa_m = \gamma = 0$, with varying $\delta_1$. 
			(c) $\delta_1 = \delta_3 = 1$, $g_a = g_b = 1$, $\kappa_a = \kappa_b = \kappa_m = \gamma = 0$, with varying $\delta_2$. 
			(d)  $\delta_1 = \delta_2 = 1$, $g_a = g_b = 1$, $\kappa_a = \kappa_b = \kappa_m = \gamma = 0$, with varying $\delta_3$. 
			(e)  $\delta_i = 1$ for $i = 1, 2, 3$, $g_b = 1$, $\kappa_a = \kappa_b = \kappa_m = \gamma = 0$, with varying $g_a$. 
			(f)  $\delta_i = 1$ for $i = 1, 2, 3$, $g_a = 1$, $\kappa_a = \kappa_b = \kappa_m = \gamma = 0$, with varying $g_b$. 
			(g)  $\delta_i = 1$ for $i = 1, 2, 3$, $g_a = g_b = 1$, $\kappa_a = \kappa_b = \kappa_m = 0$, with varying $\gamma$. 
			(h)  $\delta_i = 1$ for $i = 1, 2, 3$, $g_a = g_b = 1$, with varying $\kappa_a, \kappa_b, \kappa_m$.}
		\label{f1}
	\end{figure}
Figure \ref{f1} illustrates the temporal evolution of the charger's coherence, $C(t)$. The oscillatory behavior is a direct signature of coherent excitation exchange among the charger's constituent modes, driven by the initial energy transfer from the atomic subsystem.
	
The influence of the system's coupling strengths is depicted in Figs. \ref{f1}(a), \ref{f1}(e), and \ref{f1}(f). The atom-cavity coupling $\lambda$ initiates the process, while the internal charger couplings $g_a$ and $g_b$ mediate subsequent energy exchange. Increasing any of these parameters enhances the magnitude of the off-diagonal elements in the system's evolution matrix $\mathcal{A}$. This results in faster and more pronounced Rabi-like oscillations, leading to higher peak coherence values. This demonstrates that strong coherent driving is essential for rapidly establishing a robust superposition across the charger states.

The effect of frequency detuning is shown in Figs. \ref{f1}(b)-\ref{f1}(d). The parameters $\delta_i$ contribute to the diagonal elements of the evolution matrix, representing energy mismatches between the coupled modes. Operating off-resonance ($\delta_i \neq 0$) suppresses the efficiency of population transfer, which manifests as a significant reduction in the amplitude of coherence oscillations. Optimal coherence is therefore strictly achieved under the resonant condition $\delta_i = 0$, where energy exchange is most efficient. The impact of environmental decoherence is presented in Figs. \ref{f1}(g) and \ref{f1}(h). The atomic decay rate $\gamma$ and the charger decay rates $\kappa_{a,b,m}$ introduce imaginary components to the diagonal elements of $\mathcal{A}$. This leads to an exponential damping of the probability amplitudes $C_n(t)$. As coherence is proportional to products of these amplitudes, it is highly sensitive to dissipation. Fig. \ref{f1}(g) shows that atomic decay acts as a loss from the energy source, while Fig. \ref{f1}(h) demonstrates that direct decay from the charger modes rapidly destroys the phase relationships, driving the system towards an incoherent mixed state.

Overall, Fig. \ref{f1} provides a comprehensive map of the charger's coherence dynamics. It establishes that the generation of coherence is a coherent process driven by strong coupling, while its persistence is fundamentally limited by both energy mismatches (detuning) and irreversible environmental coupling (dissipation). The results collectively underscore that achieving and maintaining high coherence requires a finely tuned system operating at resonance with minimal decoherence.

	\section{Energy}\label{sec3}
	The stored energy in a quantum battery represents the extractable work accumulated within its quantum states \cite{zahia2024quantum,campaioli2024colloquium}. It can be expressed as  
	\begin{align}
		E(t) &= Tr(\hat{\rho}_ {QB}(t)\hat{H}_{QB}) -Tr(\hat{\rho}_ {QB}(0)\hat{H}_{QB})\notag\\
		&=\omega_{q}\left(1-|C_1(t)|^2-|C_2(t)|^2-|C_3(t)|^2  \right)
	\end{align}

	where $\hat{\rho}_ {QB}(t)$ is the battery's density matrix, $\hat{H}_{QB}=\sum_{j=1}^{2} \left(
	\frac{\omega_{q}}{2}\, \hat{\sigma}_z^{(j)} \right)$ is the Quantum Battery Hamiltonian.
	
	\begin{figure}[h!]
		\centering
		\includegraphics[width=1.05\textwidth]{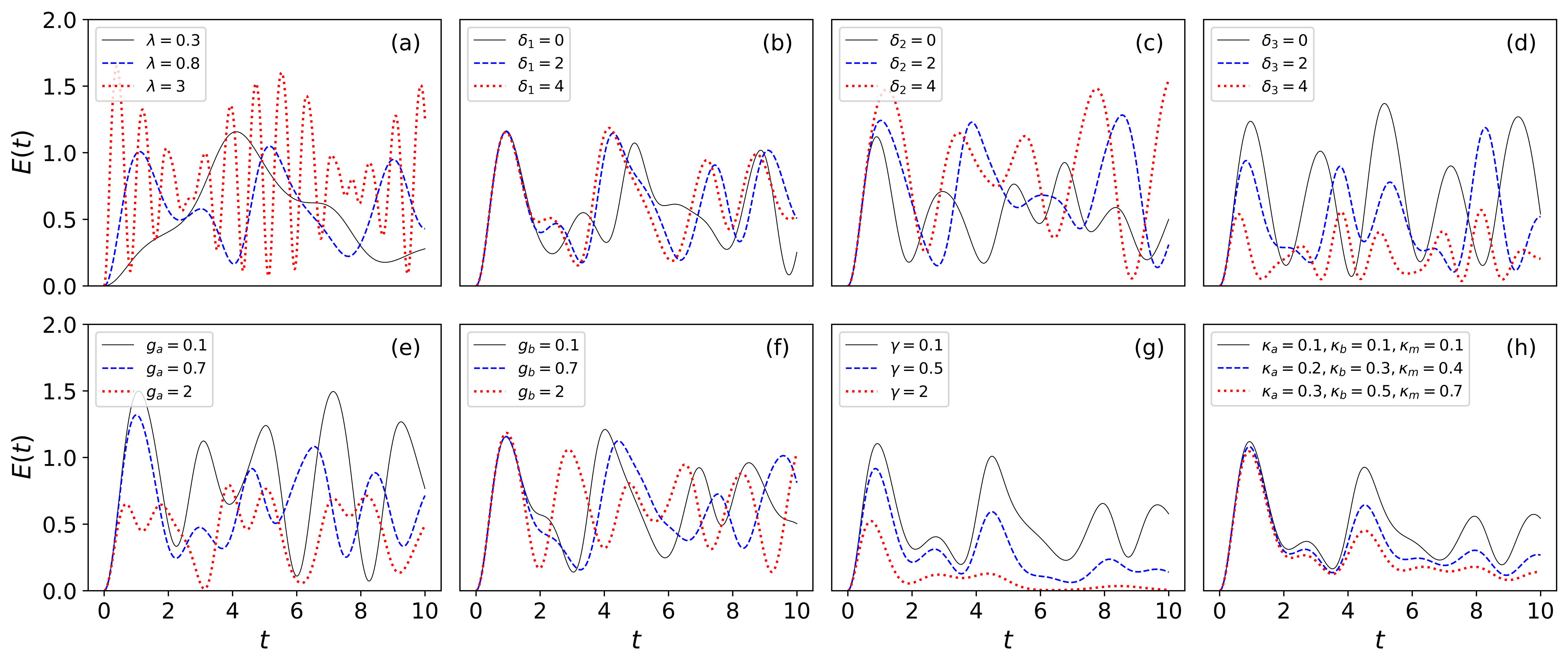}
		\caption{Energy dynamics $E(t)$ as a function of time $t$ for the same parameters as given in Figure \ref{f1}}
		\label{f2}
	\end{figure}
	
	 The dynamics of the stored energy $E(t)$ is displayed in Fig \ref{f2}, which is proportional to the excited state population of the atomic battery. The observed oscillations are a direct manifestation of the coherent and reversible energy exchange between the charger and the battery subsystems. The role of the coherent coupling parameters is examined in Figs. \ref{f2}(a), \ref{f2}(e), and \ref{f2}(f). The strengths $\lambda$, $g_a$, and $g_b$ constitute the off-diagonal elements of the evolution matrix $\mathcal{A}$ and dictate the transition rates between states. Increasing these couplings results in higher frequency oscillations and a larger peak stored energy. This occurs because a larger $\lambda$ enhances the direct energy transfer rate to the battery, while larger $g_a$ and $g_b$ accelerate the propagation of the excitation within the charger, making it more readily available for transfer. 
	 
	 The influence of detuning is shown in Figs. \ref{f2}(b)-\ref{f2}(d). The detunings $\delta_i$ appear in the diagonal elements of $\mathcal{A}$, representing energy mismatches that disrupt resonant transfer. Consequently, any non-zero detuning suppresses the amplitude of energy oscillations and reduces the maximum achievable charge. The optimal charging condition is therefore strictly limited to the resonant regime where all $\delta_i = 0$. The effects of dissipation are illustrated in Figs. \ref{f2}(g) and \ref{f2}(h). The decay rates $\gamma$ and $\kappa_{a,b,m}$ introduce imaginary components to the diagonal of $\mathcal{A}$, leading to an overall decay of the state amplitudes. Atomic decay $\gamma$ is a direct loss channel for the energy stored in the battery. In contrast, the charger decay rates $\kappa$ deplete the energy available for transfer, thereby diminishing the charging efficiency. This interplay between coherent oscillation and exponential decay underscores the importance of temporal control for extracting energy at the optimal instant.

	 Generally, Fig. \ref{f2} provides a quantitative analysis of the battery's charging cycle. It demonstrates that the charging process is a competition between coherent mechanisms that store energy and incoherent processes that dissipate it. The figure clearly establishes that maximal energy storage is achieved through a combination of strong, resonant coupling and minimal environmental decoherence, highlighting that both the rate and the efficiency of charging are critical performance metrics.

	\section{Ergotropy}\label{sec4}
	The concept of \textit{ergotropy} plays a fundamental role in assessing the efficiency of a quantum battery (QB) \cite{campaioli2024colloquium,zahia2025entanglement}. Ergotropy is defined as the maximum amount of energy stored in a quantum state that can be converted into useful (extractable) work. It is quantified as the difference between the internal energy of the quantum state $\hat{\rho}_B$ (representing the atomic state of the battery) and the internal energy of its corresponding \textit{passive state} $\hat{\eta}$, which is a state from which no work can be extracted under unitary operations. Mathematically, this is expressed as:
	\begin{equation}
		\epsilon(\hat{\rho}_{battery}, \hat{H}_{QB}) = E(\hat{\rho}_{battery}, \hat{H}_{QB}) - E(\hat{\eta}, \hat{H}_{QB}),
	\end{equation}
	
	where the energy is computed as $E(\hat{\rho}_{battery}, \hat{H}_{QB}) = \text{Tr}(\hat{\rho}_{battery} \hat{H}_{QB})$. The passive state $\hat{\eta}$ is defined as a state that is diagonal in the eigenbasis of the Hamiltonian $\hat{H}_{QB}$ and exhibits a non-increasing population with respect to increasing energy levels. Furthermore, the passive state commutes with the Hamiltonian, i.e., $[\hat{H}_{QB}, \hat{\eta}] = 0$.
	
	Let the spectral decompositions of the battery state and its Hamiltonian be given by
	\[
	\hat{\rho}_{battery} = \sum_{i=1}^{d} p_i \ket{p_i} \bra{p_i}, \quad \hat{H}_{QB} = \sum_{i=1}^{d} \epsilon_i \ket{\epsilon_i} \bra{\epsilon_i},
	\]
	where the eigenvalues satisfy $p_1 \geq p_2 \geq \cdots \geq p_d$ and $\epsilon_1 \leq \epsilon_2 \leq \cdots \leq \epsilon_d$, with $d$ denoting the dimension of the system’s Hilbert space. Under this ordering, the corresponding passive state is constructed as:
	
	\[
	\hat{\eta} = \sum_{i=1}^{d} p_i \ket{\epsilon_i} \bra{\epsilon_i}.
	\]
	
	Accordingly, the ergotropy can be expressed in terms of the overlap between the energy and state eigenbases as:
	
	\begin{equation}
		\epsilon(\hat{\rho}_{battery}, \hat{H}_{QB}) = \sum_{i,j=1}^{d} p_i \epsilon_j \left( |\braket{p_i | \epsilon_j}|^2 - \delta_{ij} \right),
	\end{equation}
	
	where $\delta_{ij}$ is the Kronecker delta. 
	\begin{figure}[h!]
		\centering
		\includegraphics[width=1.05\textwidth]{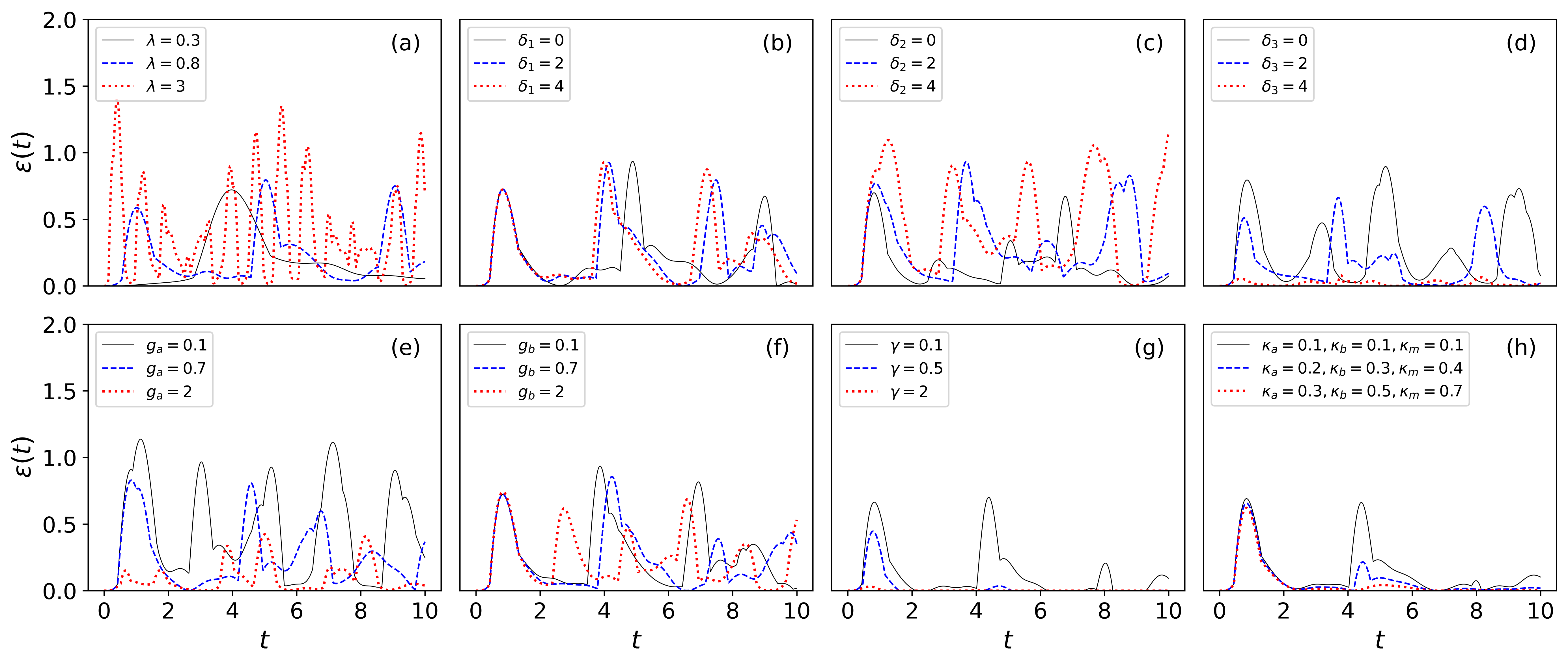}
		\caption{Ergotropy dynamics $\epsilon(t)$ as a function of  time $t$ for the same parameters as shown in Figure \ref{f1}}
		\label{f3}
	\end{figure}

Figure \ref{f3} presents the ergotropy $\mathcal{E}(t)$, which quantifies the maximum extractable work from the battery. Ergotropy is a stricter performance metric than stored energy, as it depends not only on the battery's state populations but also on its internal coherence. A comparison with Fig. \ref{f2} confirms the thermodynamic bound $\mathcal{E}(t) \le E(t)$, where the difference $E(t) - \mathcal{E}(t)$ represents the passive energy locked in incoherent state components.

The dynamics of ergotropy under unitary evolution are shown in Figs. \ref{f3}(a)-\ref{f3}(f). The behavior mirrors that of the stored energy in Fig. \ref{f2}, with strong couplings ($\lambda, g_a, g_b$) and resonant conditions ($\delta_i=0$) maximizing the peak values. This indicates that under ideal conditions, the charging process is highly coherent, and the stored energy is almost entirely extractable as work. The primary role of the charger's internal coherence, analyzed in Fig. \ref{f1}, is to facilitate this efficient and coherent transfer of energy to the battery. The effects of dissipation are shown in Figs. \ref{f3}(g) and \ref{f3}(h). A crucial distinction emerges when comparing these to the energy dynamics in Fig. \ref{f2}. The decay of ergotropy is significantly more rapid than the decay of total energy. This is because dissipation attacks the system on two fronts. It reduces the excited state population, which lowers $E(t)$, and it destroys the off-diagonal elements of the battery's density matrix, which directly converts extractable work into passive energy. Ergotropy is therefore more sensitive to the decoherence channels governed by $\gamma$ and $\kappa_{a,b,m}$ than the total stored energy is.

In summary, Fig \ref{f3} provides a nuanced view of the battery's functional performance. By comparing it with the previous figures, it becomes clear that while strong charger coherence (Fig. \ref{f1}) and high stored energy (Fig. \ref{f2}) are necessary conditions for good performance, they are not sufficient. The ergotropy analysis reveals that preserving the battery's internal coherence against dissipation is the most critical factor for maximizing the truly useful work that can be extracted, making it the most definitive metric for battery efficiency.

\begin{figure}[h!]
	\centering

	\subfloat[]{%
		\includegraphics[width=0.47\textwidth]{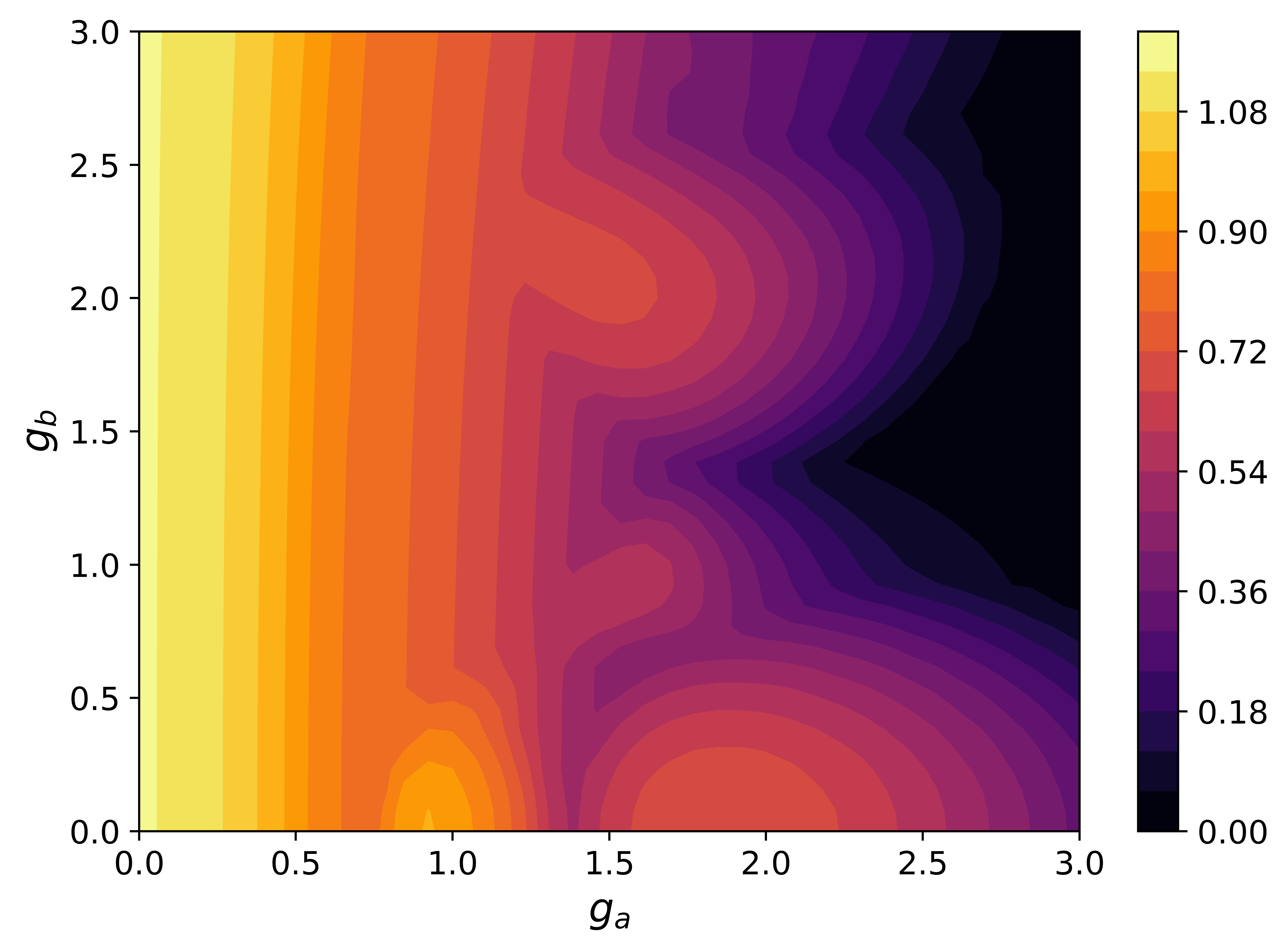}
	}
	\hspace{0\textwidth}
	\subfloat[]{%
		\includegraphics[width=0.47\textwidth]{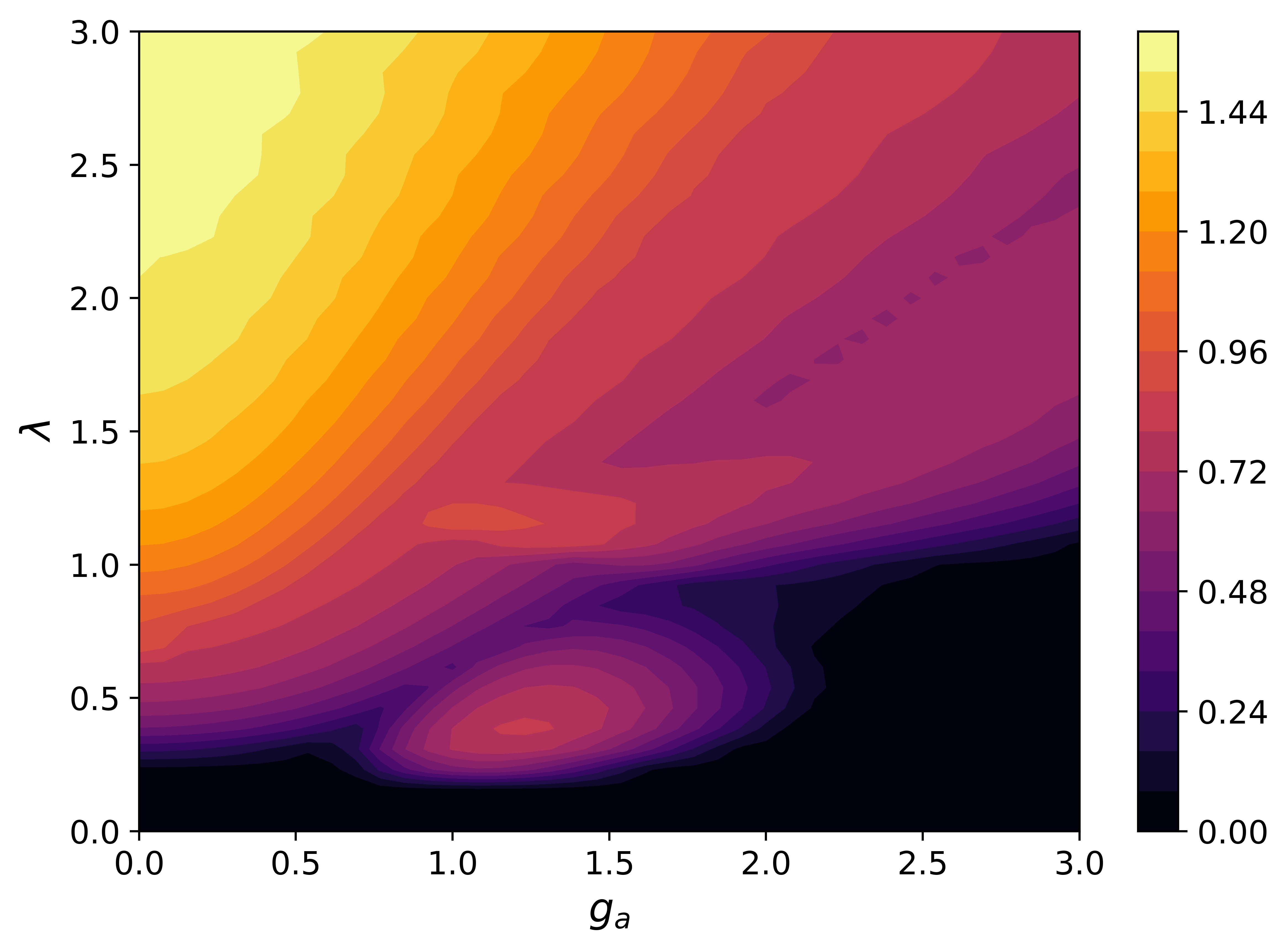}
	}
	\vspace{0\textwidth}
	\subfloat[]{%
		\includegraphics[width=0.47\textwidth]{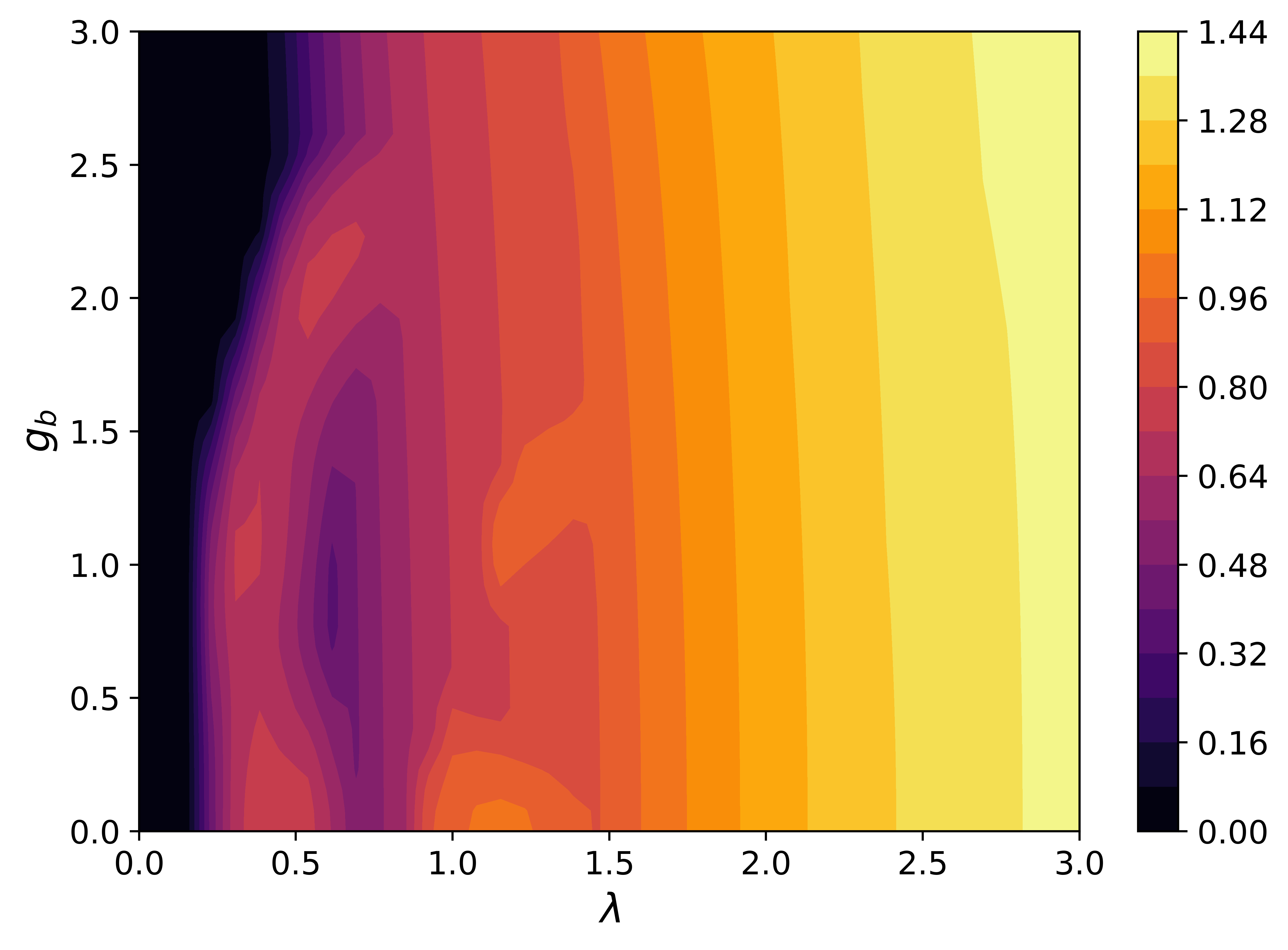}
	}
	\caption{Contour plots of the maximum ergotropy as functions of the coupling parameters. We set $\delta_i = 1$ for $i = 1,2,3$, and $\kappa_a = \kappa_b = \kappa_m = \gamma = 0$. (a) $g_a$ versus $g_b$ with $\lambda = 1$. (b) $g_a$ versus $\lambda$ with $g_b = 1$. (c) $\lambda$ versus $g_b$ with $g_a = 1$.}
	\label{f4}
\end{figure}
To gain deeper insight into the interplay of coherent interactions, the contour plots in Fig. \ref{f4} map the maximum achievable ergotropy, $\mathcal{E}_{\text{max}}$, across the parameter space of the coupling strengths $\lambda$, $g_a$, and $g_b$. These plots complement the temporal analysis by identifying the optimal operating points for peak work extraction. A key feature revealed in all three figures is the non-monotonic dependence of $\mathcal{E}_{\text{max}}$ on the coupling strengths. This behavior arises from the hybridization of the bare energy states into the dressed states (or normal modes) of the coupled system. The energy exchange dynamics are governed by the interference between these dressed states. While stronger coupling generally increases the energy exchange rate, excessively strong coupling in one channel relative to others can lead to destructive interference or impedance mismatch. For instance, Fig. \ref{f4}(a) shows that for fixed $\lambda$, an intermediate value of $g_a$ is optimal. A very large $g_a$ creates strongly hybridized cavity-magnon polaritons, which may couple less efficiently to the atomic subsystem, thus hindering the overall energy transfer to the battery. Figure \ref{f4}(b) illustrates the cooperative yet competitive relationship between $\lambda$ and $g_a$. A larger $\lambda$ is required to initiate a faster transfer, but an excessive $g_a$ can effectively trap the excitation within the charger's cavity-magnon subsystem, preventing its efficient delivery to the atoms. Similarly, Fig. \ref{f4}(c) shows that the influence of the magnon-phonon coupling $g_b$ is amplified by a strong atom-cavity coupling $\lambda$. This highlights that the entire chain of interactions from atom to cavity to magnon to phonon must be well-balanced to create a clear and efficient pathway for the excitation to travel.

\begin{figure}[h!]
	\centering
	
	\subfloat[]{%
		\includegraphics[width=0.47\textwidth]{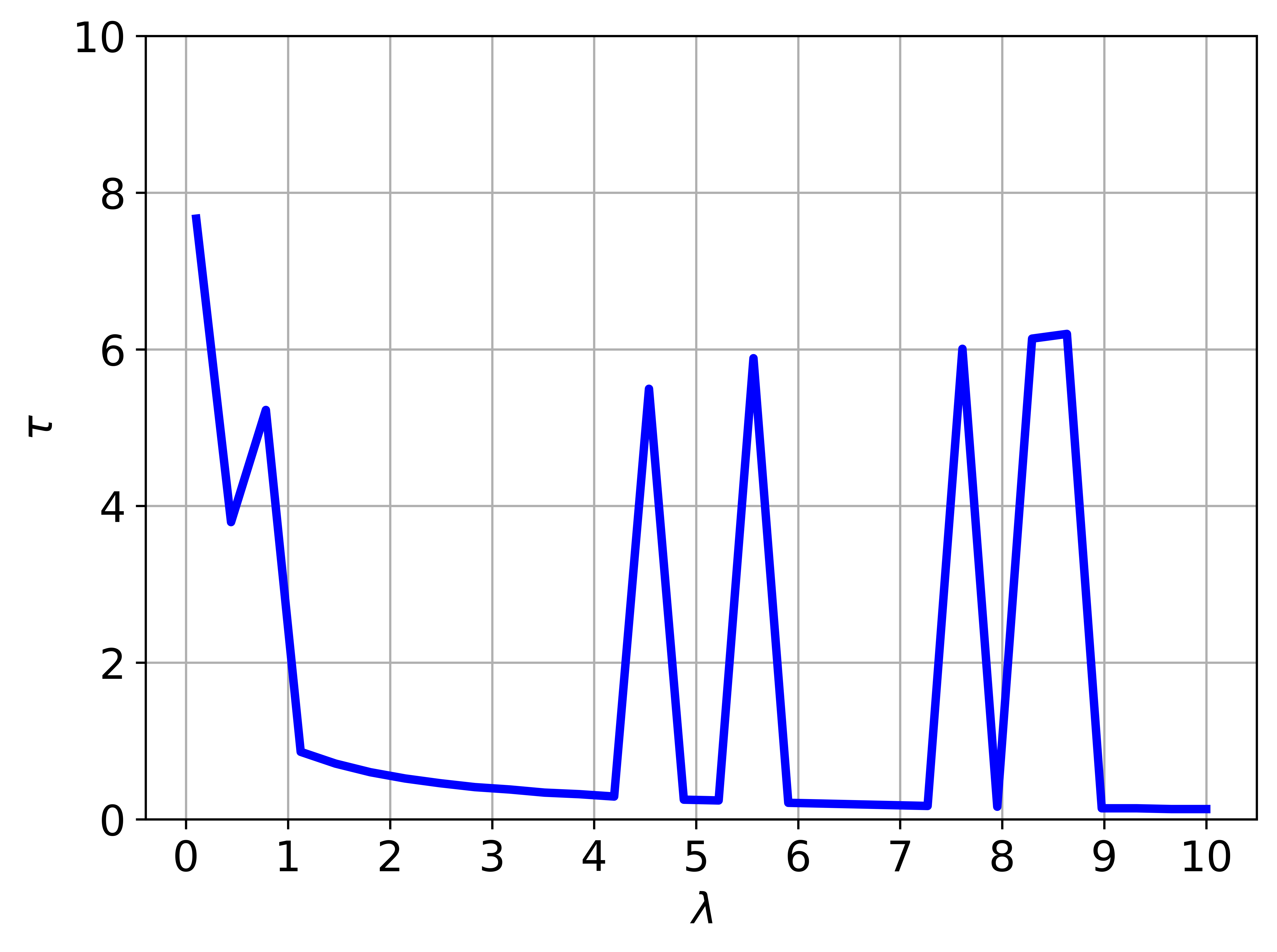}
	}
	\hspace{0\textwidth}
	\subfloat[]{%
		\includegraphics[width=0.47\textwidth]{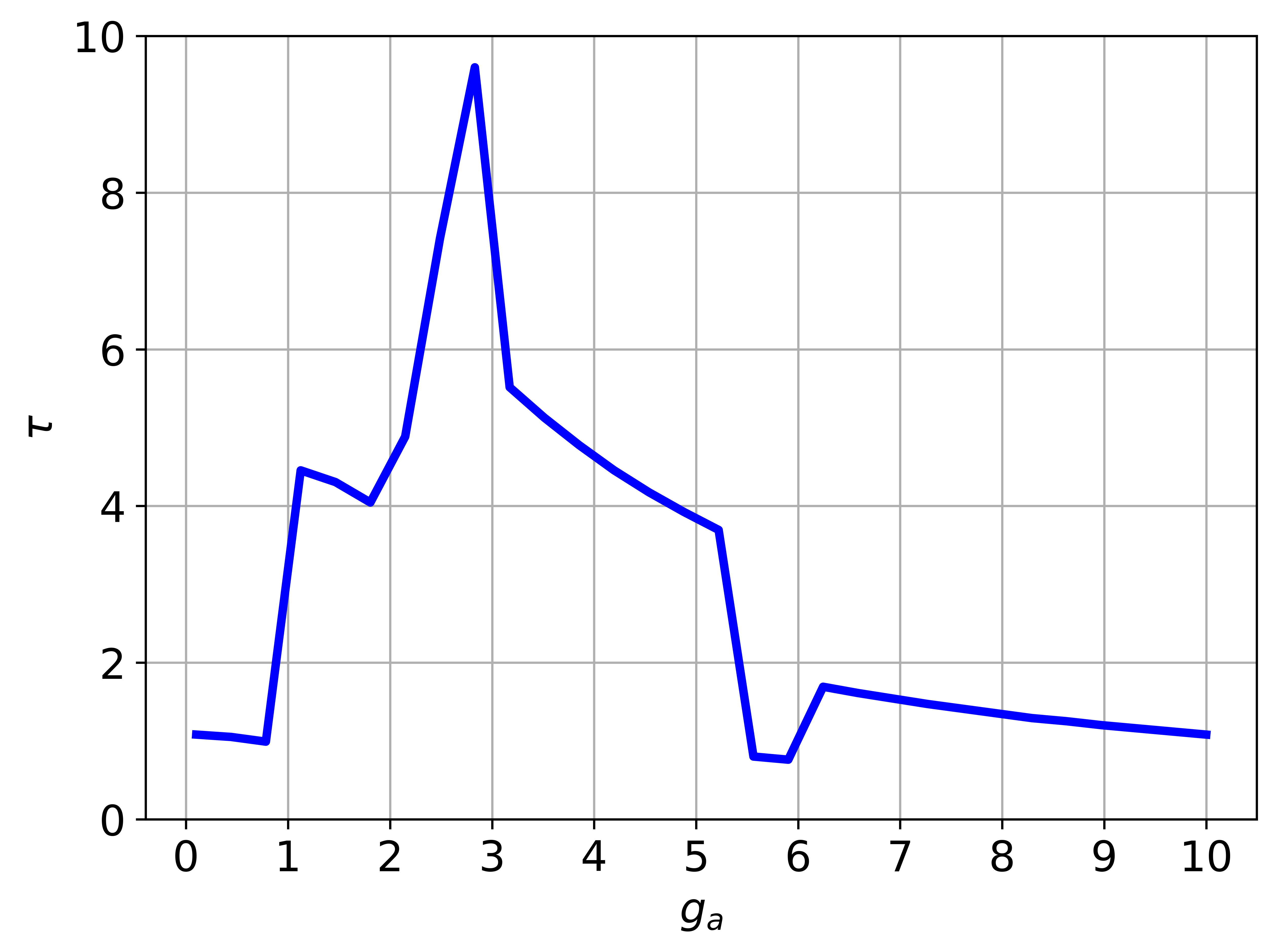}
	}
	\vspace{0\textwidth}
	\subfloat[]{%
		\includegraphics[width=0.47\textwidth]{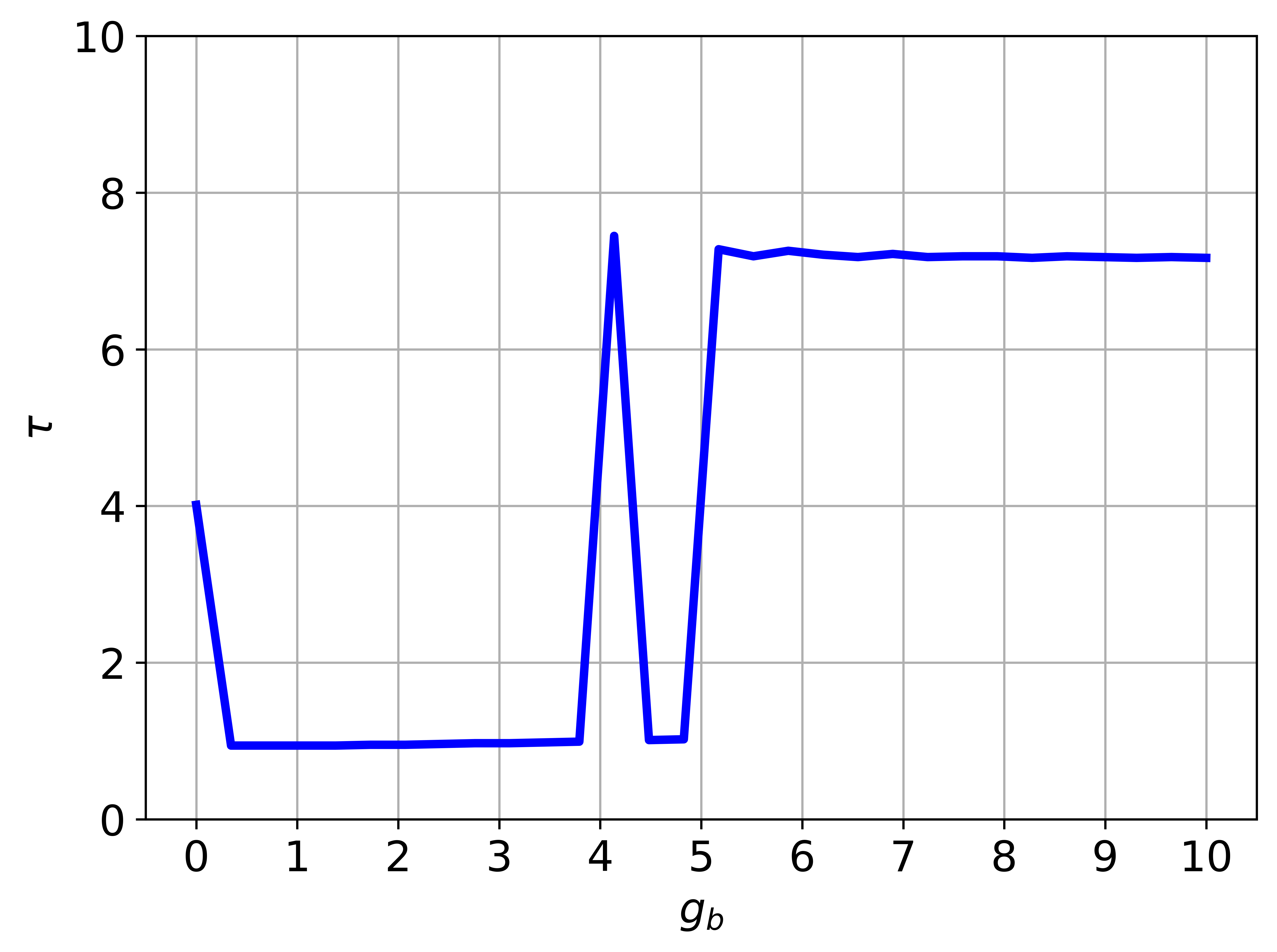}
	}
	\caption{Optimal charging time $\tau$ for achieving maximum stored energy as a function of the coupling parameters. We set $\delta_i = 1$ for $i = 1,2,3$ and $\kappa_a = \kappa_b = \kappa_m = \gamma = 0$. (a) Optimal time $\tau$ for $g_a = g_b = 1$, with $\lambda$ varied. (b) Optimal time $\tau$ for $\lambda = g_b = 1$, with $g_a$ varied. (c) Optimal time $\tau$ for $\lambda = g_a = 1$, with $g_b$ varied.}
\label{f5}	
\end{figure}

The optimal charging time $\tau$, corresponding to the maximum stored energy, exhibits a nontrivial dependence on the system’s coupling parameters, as shown in Fig.~\ref{f5}. The highly irregular behavior of $\tau$ seen in Figs. \ref{f5}(a) and \ref{f5}(b) is a direct consequence of the multi-level interference discussed for Fig. 5. The charging time is determined by the generalized Rabi frequencies of the system, which are functions of the eigenvalues of the evolution matrix $\mathcal{A}$. As the coupling parameters are varied, the dressed-state energy levels undergo avoided crossings. Near these crossings, the character of the eigenstates changes abruptly, leading to constructive or destructive interference that can either accelerate or significantly slow down the population transfer to the battery, resulting in the sharp peaks and troughs observed in $\tau$.

A comparison with Fig. \ref{f4} is particularly insightful. The regions in the parameter space of Fig. \ref{f4} that yield the highest maximum ergotropy do not always correspond to the shortest charging times in Fig. \ref{f5}. For instance, a parameter set might enable a very high-fidelity energy transfer (high $\mathcal{E}_{\text{max}}$) but over a longer period (large $\tau$). This reveals a fundamental trade-off between charging efficiency (the fraction of energy that is extractable) and charging power (the speed of charging). Fig. \ref{f5}(c) shows that beyond a certain coupling strength $g_b$, the charging time saturates. This suggests a regime where the magnon-phonon interaction is no longer the rate-limiting step in the energy transfer process.

\section{Purity}\label{pur}
Purity quantifies the degree of coherence in a quantum state and distinguishes pure states from mixed states. For quantum batteries, maintaining high purity ensures efficient charging dynamics, maximizes extractable energy, and enhances overall performance \cite{arjmandi2022performance}.
	\begin{align}
		\mathcal{P}(t)=Tr[\hat{\rho}_{\mathrm{QB}}(t)\hat{\rho}_{\mathrm{QB}}(t)] =\ & A^2 \ket{gg}\bra{gg}
		+ B^2 \ket{eg}\bra{eg}
		+ (C^2+DE) \ket{ge}\bra{ge} \notag \\
		& + (BE+EC) \ket{eg}\bra{ge}
		+ (CD+BD) \ket{ge}\bra{eg}.
 	\end{align}
where $A=\left(|C_1(t)|^2 + |C_2(t)|^2 + |C_3(t)|^2\right)$, $B=|C_4(t)|^2$, $C=|C_5(t)|^2$, $D=C_4(t)C_5^*(t)$, and $D=C_4^*(t)C_5(t)$
	
		\begin{figure}[h!]
		\centering
		\includegraphics[width=1.05\textwidth]{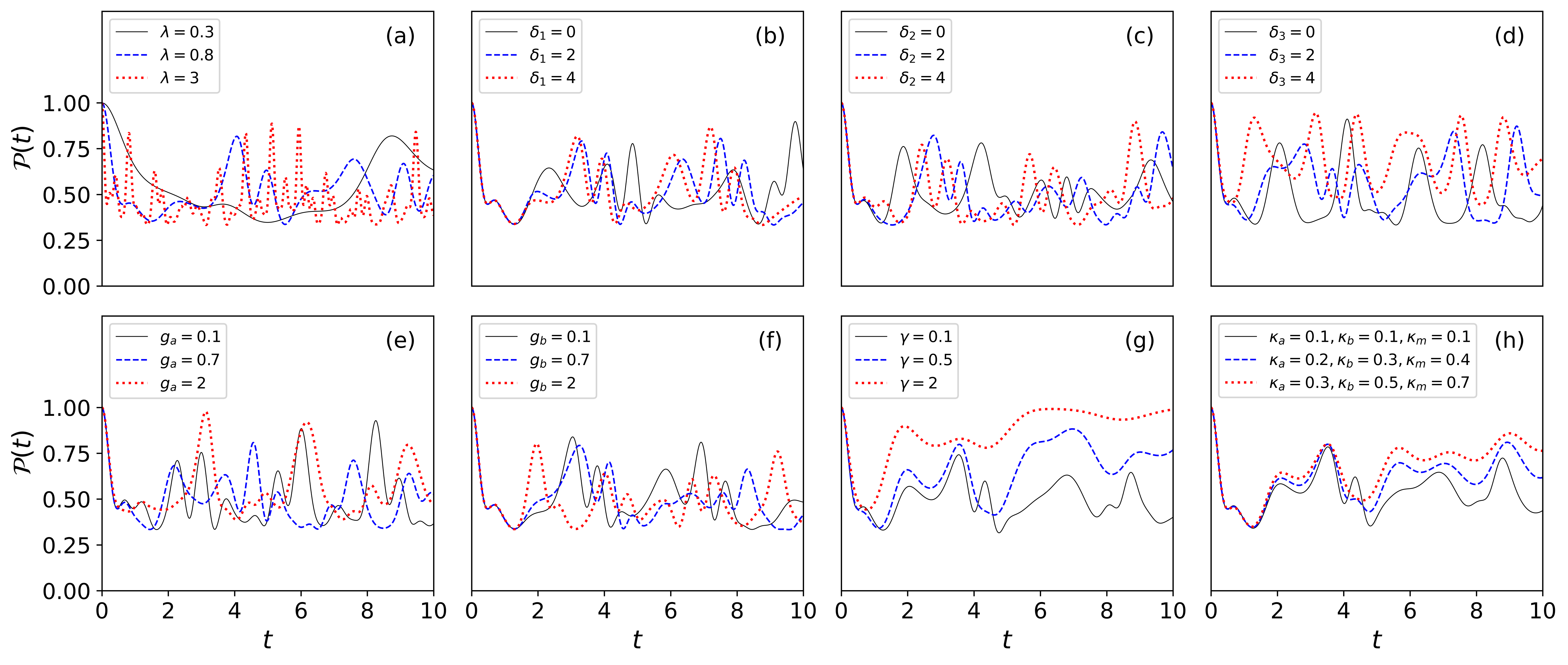}
		\caption{Purity dynamics $\mathcal{P}(t)$ as a function of  time $t$ for the same parameters as shown in Figure \ref{f1}}
		\label{f6}
	\end{figure}
Temporal evolution of the purity, $\mathcal{P}(t)$, provides key insights into the coherence dynamics of the quantum battery during the charging process, as shown in Fig.~\ref{f6}.

First, we compare the unitary dynamics shown in Figs. \ref{f6}(a)-\ref{f6}(f) with the charger coherence in Fig. 2 and the stored energy in Fig. 3. The oscillations in the battery's purity are a direct signature of coherent information exchange. A distinct anti-correlation exists between the battery's purity $\mathcal{P}(t)$ and the charger's coherence $C(t)$. When the charger's modes are in a robust superposition (high $C(t)$), the excitation is delocalized across the charger, resulting in a strong entanglement with the battery. This entanglement necessarily implies that the battery's reduced state is mixed, hence its purity $\mathcal{P}(t)$ is low. Conversely, when the energy is maximally transferred to the battery (peaks in Fig. \ref{f2}), the battery-charger system becomes disentangled, and the battery ideally occupies a pure excited state, causing the purity $\mathcal{P}(t)$ to approach unity. Therefore, the oscillations in purity represent the periodic swapping of quantum coherence between the battery and charger subsystems.

The impact of dissipation is shown in Figs. \ref{f6}(g) and \ref{f6}(h), which provided the definitive physical explanation for the divergence between stored energy (Fig. \ref{f2}) and ergotropy (Fig. \ref{f3}). The decay of purity under dissipation is irreversible and is mathematically more severe than the decay of energy. Stored energy $E(t)$ is a function of the diagonal elements of $\hat{\rho}_{\text{QB}}$ (populations), which decay at rates related to $\gamma$. In contrast, ergotropy $\mathcal{E}(t)$ and purity $\mathcal{P}(t)$ are highly sensitive to the off-diagonal elements of $\hat{\rho}_{\text{QB}}$ (coherences), which are destroyed by all decoherence mechanisms. The rapid, irreversible decay of $\mathcal{P}(t)$ signifies the environment effectively "measuring" the battery, collapsing its superposition and projecting it into a classical statistical mixture. This very process of destroying quantum coherence is the mechanism that converts the ordered, ergotropy into disordered, passive energy. Thus, the faster decay of ergotropy compared to total energy is a direct consequence of the rapid decay of purity.

\section{Conclusion}\label{sec5}
In this work, we have theoretically investigated a hybrid cavity magnomechanical system as a platform for realizing a quantum battery. The model incorporates two identical two-level atoms coupled to a single mode microwave cavity and a YIG sphere based magnon mode, enabling coherent photon magnon atom interactions for efficient quantum energy storage. Within the rotating-wave approximation, we derived and solved the system Hamiltonian to investigate the temporal evolution of the quantum battery in terms of coherence, stored energy, and ergotropy. Our results show that strong microwave cavity magnon coupling significantly enhances both the charging efficiency and the extractable work, whereas various detuning and dissipation mechanisms such as atomic decay, photon- magnon-phonon losses degrade performance by suppressing coherence and energy retention. These findings emphasize the importance of maintaining on resonant coupling and minimizing different loss channels to achieve stable and efficient operation. Overall, the study elucidates the underlying mechanisms governing the performance of hybrid cavity magnomechanical framework for quantum batteries, offering practical insights for optimizing their design and advancing the realization of high-efficiency quantum energy storage in emerging quantum technologies.	
	
\section*{Author Contributions Statement}
S.~K.~Singh and J.X. Peng conceptualise the idea and prepared the introduction section. Ahmed A.~Zahia carried out the analytical calculations and numerical simulations. M.~Y.~Abd-Rabbou, S.K.Singh and J.X. Peng drafted the academic discussion and revised the entire manuscript. 
\section*{Ethics declarations}
The authors declare no competing interests.

\section*{Availability of data and materials}
The used code of this study is available from the corresponding author upon reasonable
request.

\section*{Acknowledgment}
Jia-Xin Peng is supported by the National Natural Science Foundation of China (Grant No.~12504566), 
the Natural Science Foundation of Jiangsu Province (Grant No.~BK20250947), 
the Natural Science Foundation of the Jiangsu Higher Education Institutions (Grant No.~25KJB140013), 
and the Natural Science Foundation of Nantong City (Grant No.~JC2024045). 
SKS gratefully acknowledges the High Performance Computing facilities provided by Akal University, Punjab, India.

\bibliographystyle{unsrt}
\bibliography{sample}

\end{document}